\documentclass[12pt,epsf]{article}
\usepackage{graphicx,amsmath,amssymb}
\begin{document}

\def\a{\alpha}
\def\b{\beta}
\def\c{\varepsilon}
\def\d{\delta}
\def\e{\epsilon}
\def\f{\phi}
\def\g{\gamma}
\def\h{\theta}
\def\k{\kappa}
\def\l{\lambda}
\def\m{\mu}
\def\n{\nu}
\def\p{\psi}
\def\q{\partial}
\def\r{\rho}
\def\s{\sigma}
\def\t{\tau}
\def\u{\upsilon}
\def\v{\varphi}
\def\w{\omega}
\def\x{\xi}
\def\y{\eta}
\def\z{\zeta}
\def\D{\Delta}
\def\G{\Gamma}
\def\H{\Theta}
\def\L{\Lambda}
\def\F{\Phi}
\def\P{\Psi}
\def\S{\Sigma}

\def\o{\over}
\def\beq{\begin{eqnarray}}
\def\eeq{\end{eqnarray}}
\newcommand{\gsim}{ \mathop{}_{\textstyle \sim}^{\textstyle >} }
\newcommand{\lsim}{ \mathop{}_{\textstyle \sim}^{\textstyle <} }
\newcommand{\vev}[1]{ \left\langle {#1} \right\rangle }
\newcommand{\bra}[1]{ \langle {#1} | }
\newcommand{\ket}[1]{ | {#1} \rangle }
\newcommand{\EV}{ {\rm eV} }
\newcommand{\KEV}{ {\rm keV} }
\newcommand{\MEV}{ {\rm MeV} }
\newcommand{\GEV}{ {\rm GeV} }
\newcommand{\TEV}{ {\rm TeV} }
\def\diag{\mathop{\rm diag}\nolimits}
\def\Spin{\mathop{\rm Spin}}
\def\SO{\mathop{\rm SO}}
\def\O{\mathop{\rm O}}
\def\SU{\mathop{\rm SU}}
\def\U{\mathop{\rm U}}
\def\Sp{\mathop{\rm Sp}}
\def\SL{\mathop{\rm SL}}
\def\tr{\mathop{\rm tr}}

\def\IJMP{Int.~J.~Mod.~Phys. }
\def\MPL{Mod.~Phys.~Lett. }
\def\NP{Nucl.~Phys. }
\def\PL{Phys.~Lett. }
\def\PR{Phys.~Rev. }
\def\PRL{Phys.~Rev.~Lett. }
\def\PTP{Prog.~Theor.~Phys. }
\def\ZP{Z.~Phys. }

\newcommand{\bea}{\begin{eqnarray}}   
\newcommand{\eea}{\end{eqnarray}}
\newcommand{\bear}{\begin{array}}  
\newcommand {\eear}{\end{array}}
\newcommand{\bef}{\begin{figure}}  
\newcommand {\eef}{\end{figure}}
\newcommand{\bec}{\begin{center}}  
\newcommand {\eec}{\end{center}}
\newcommand{\non}{\nonumber}  
\newcommand {\eqn}[1]{\beq {#1}\eeq}
\newcommand{\la}{\left\langle}  
\newcommand{\ra}{\right\rangle}
\newcommand{\ds}{\displaystyle}
\def\SEC#1{Sec.~\ref{#1}}
\def\FIG#1{Fig.~\ref{#1}}
\def\EQ#1{Eq.~(\ref{#1})}
\def\EQS#1{Eqs.~(\ref{#1})}
\def\TEV#1{10^{#1}{\rm\,TeV}}
\def\GEV#1{10^{#1}{\rm\,GeV}}
\def\MEV#1{10^{#1}{\rm\,MeV}}
\def\KEV#1{10^{#1}{\rm\,keV}}
\def\lrf#1#2{ \left(\frac{#1}{#2}\right)}
\def\lrfp#1#2#3{ \left(\frac{#1}{#2} \right)^{#3}}
\def\REF#1{Ref.~\cite{#1}}
\newcommand{\osc}{{\rm osc}}
\newcommand{\ed}{{\rm end}}
\def\dda#1{\frac{\partial}{\partial a_{#1}}}
\def\ddat#1{\frac{\partial^2}{\partial a_{#1}^2}}
\def\dd#1#2{\frac{\partial #1}{\partial #2}}
\def\ddt#1#2{\frac{\partial^2 #1}{\partial #2^2}}
\def\lrp#1#2{\left( #1 \right)^{#2}}


\baselineskip 0.7cm

\begin{titlepage}

\begin{flushright}
TU-959\\
IPMU14-0059\\
\end{flushright}

\vskip 1.35cm
\begin{center}
{\large \bf 
Higgs Chaotic Inflation and the Primordial B-mode Polarization
Discovered by BICEP2
}
\vskip 1.2cm
Kazunori Nakayama$^{a,c}$,
Fuminobu Takahashi$^{b,c}$

\vskip 0.4cm

{\it $^a$Department of Physics, University of Tokyo, Tokyo 113-0033, Japan}\\
{\it $^b$Department of Physics, Tohoku University, Sendai 980-8578, Japan}\\
{\it $^c$Kavli Institute for the Physics and Mathematics of the Universe (WPI), TODIAS, University of Tokyo, Kashiwa 277-8583, Japan}

\vskip 1.5cm

\abstract{
We show that the standard model Higgs field can realize the quadratic chaotic inflation, 
if the kinetic term is significantly modified at large field values. This is a simple realization of the so-called running kinetic inflation. The point is that the Higgs field respects an approximate shift symmetry at high energy scale.
The  tensor-to-scalar ratio is predicted to be $r \simeq 0.13 - 0.16$, which
nicely explains the primordial B-mode polarization, $r=0.20^{+0.07}_{-0.05}$,  recently discovered by the BICEP2 experiment. 
In particular, allowing small modulations induced by the shift symmetry breaking, the negative running spectral index can also be
induced.  The reheating temperature is expected to be so high that successful thermal 
leptogenesis is possible. 
The suppressed quartic coupling of the Higgs field at high energy scales may be related to the Higgs chaotic inflation. 
}
\end{center}
\end{titlepage}

\setcounter{page}{2}

Our Universe experienced an accelerated expansion at a very early stage
of the evolution, i.e., inflation~\cite{Guth:1980zm,Linde:1981mu}. 
Among various inflation models proposed so far, 
the so-called chaotic inflation~\cite{Linde:1983gd} is particularly
interesting as it predicts large values of the tensor-to-scalar ratio $r$.
The tensor mode density perturbations generate the primordial B-mode polarization of
the cosmic microwave background (CMB),   which, if observed, would determine the
inflation scale and pin down the underlying model of inflation.

Recently the BICEP2 experiment announced that they discovered the primordial 
B-mode polarization of CMB. In terms of the tensor-to-scalar
ratio, the allowed range is given by~\cite{BICEP2}
\bea
r=0.20^{+0.07}_{-0.05} ~~(68\%{\rm CL}).
\label{B}
\eea
After subtracting the best available estimate for foreground dust, the allowed range is modified to $r = 0.16^{+0.06}_{-0.05}$.
The discovery of $r$ in this range is of significant importance for cosmology as well as particle physics, as it implies that we obtain 
the invaluable information on the Universe at the GUT scale. 

There are various large-field inflation models which predict $r$ in the above range\footnote{ 
For various large-field inflation models and their concrete realization in supergravity and superstring theory,
see e.g.~\cite{Freese:1990ni,Kawasaki:2000yn,Destri:2007pv,Silverstein:2008sg,McAllister:2008hb,Takahashi:2010ky,
Nakayama:2010kt,Nakayama:2010sk,Harigaya:2012pg,Croon:2013ana,Nakayama:2013jka,Czerny:2014wza}.}, and
 by far the simplest one is the 
quadratic chaotic inflation~\cite{Linde:1983gd}:
\bea
{\cal L} = \frac{1}{2} (\partial \phi)^2 - \frac{1}{2} m^2 \phi^2,
\eea
where $\phi$ is the inflaton and $m$ is the inflaton mass. The Planck normalization on the primordial density
perturbations~\cite{Ade:2013rta} fixes the inflaton mass as
\bea
m &\simeq& 1.5 \times \GEV{13}.
\label{m}
\eea
Since the primordial B-mode polarization was discovered, 
the next question will be the identity of the inflaton.

In fact, there is a  unique scalar field in the standard model (SM), i.e., the Higgs field, which was
discovered at LHC in 2012~\cite{Aad:2012tfa,Chatrchyan:2012ufa}.\footnote{
The connection between the SM Higgs field and inflation has been extensively 
discussed especially in a context of the non-minimal coupling to gravity~\cite{Salopek:1988qh,Bezrukov:2007ep,Einhorn:2009bh,Ferrara:2010yw,Germani:2010gm,Kamada:2010qe}.
The Starobinsky-type inflation, however, leads to much smaller values of $r$.}  
In order for the Higgs field to realize the quadratic chaotic inflation with correct density perturbations, however,
the Higgs field must have a mass of order $\GEV{13}$, many orders of magnitude larger than the observed mass, 
$m_h \approx 126$\,GeV. Moreover, the Higgs potential is dominated by the quartic term at large field values, not the quadratic one. 
The apparent discrepancy can be reconciled if either the kinetic term or the potential term is modified at large
field values.  Actually, the present authors proposed in Ref.~\cite{Nakayama:2010sk} a Higgs chaotic inflation model
in which the SM Higgs field realizes the quadratic chaotic inflation model, based on the so-called
running kinetic inflation~\cite{Takahashi:2010ky,Nakayama:2010kt}. 
In this letter we revisit the SM Higgs chaotic inflation model, in light of the
recent discoveries of the SM Higgs boson as well as the primordial CMB B-mode polarization.

The basic idea of the running kinetic inflation is very simple.  Let us consider a scalar field
with the following Lagrangian,
\bea
\label{example}
{\cal L} &=& \frac{1}{2} \left(1+ \xi \phi^2 \right) (\partial \phi)^2 - V(\phi),
\eea
where $\xi$ is a positive numerical coefficient much larger than unity, and $V(\phi)$ is the inflaton 
potential. Here and in what follows we adopt the Planck units
in which the reduced Planck scale $M_P\simeq 2.4\times 10^{18}\,$GeV is set to be unity.
Due to the dependence of the kinetic term on $\phi^2$, the canonically normalized field
at  $\phi \gtrsim 1/\sqrt{\xi}$ is given by ${\hat \phi} \sim \sqrt{\xi} \phi^2$.
As the kinetic term grows, the potential in terms of the canonically normalized field becomes
flatter. For instance, the quartic potential, $V(\phi) \sim \phi^4$, becomes the quadratic one, 
$V({\hat \phi}) \sim {\hat \phi}^2/\xi$,  at large field values. This is the essence of the running kinetic inflation.
The running kinetic inflation can be easily implemented in
 supergravity and the cosmological implications were studied in Refs.~\cite{Takahashi:2010ky,Nakayama:2010kt}.

The above argument can be straightforwardly applied to the SM Higgs field, and the SM Higgs can
drive quadratic chaotic inflation~\cite{Nakayama:2010sk}. In order to build sensible inflation models, we need to have a good control of 
the scalar potential over large field values. Also, it  is desirable to understand the large value of  $\xi \gg 1$ 
in terms of symmetry. 
To this end,  we impose an approximate shift symmetry on absolute square of the SM Higgs field $H$~\cite{Takahashi:2010ky,Nakayama:2010kt,Nakayama:2010sk}:
\beq
	|H|^2 \to |H|^2 + C,
	\label{ss}
\eeq
where $C$ is a real parameter and the SU(2)$_L$ indices are suppressed. 
We assume that the shift symmetry exhibits itself at high energy scales, whereas it is explicitly broken and therefore 
becomes much less prominent at low energy scales.
Then we can write down the Lagrangian of the Higgs field at high energy scales as follows:\footnote{
In general, we can add an arbitrary Lorentz-invariant function of  $\partial |H|^2$, which
preserves the shift symmetry, as well as other interactions of the Higgs fields,  which break the shift symmetry.
}
\beq
	\mathcal L =  \frac{1}{2} \left(\partial_\mu |H|^2\right)^2+ \epsilon |D_\mu H|^2  - \lambda \left(|H|^2 - \frac{v^2}{2}\right )^2 + \cdots,
	\label{L}
\eeq
where $\epsilon$ and $\lambda$ are coupling constants, and $D_\mu$ denotes the covariant derivative. 
The first term in (\ref{L}) respects the shift symmetry, which is explicitly broken by the second and third terms, and so,
we  expect $\epsilon, \lambda \ll 1$. Note that the second term provides the usual kinetic term for the SM Higgs
in the low energy, whereas the first term provides the kinetic term for $|H|^2$ at large field values.
In the unitary gauge, the relevant interactions are\footnote{
The Higgs chaotic inflation with the same Lagrangian was also studied in Ref.~\cite{Hertzberg:2011rc}.
}
\beq
	\mathcal L = \frac{1}{2}\left(\epsilon + h^2 \right) (\partial h)^2 - \frac{\lambda}{4}\left(h^2- v^2\right)^2,
\eeq
where $h$ denotes the physical Higgs boson, and we have omitted the gauge and Yukawa interactions which are irrelevant
during inflation. 
Thus, the largeness of $\xi$ in \EQ{example} is due to the smallness of $\epsilon$, i.e., the fact that the usual kinetic term
breaks the shift symmetry (\ref{ss}).

For large field value $h \gg \sqrt{\epsilon}$, we can rewrite the Lagrangian in terms of
the canonically normalized field $\hat h \equiv h^2/2$ as
\beq
	\mathcal L \simeq \frac{1}{2}(\partial \hat h)^2 - \lambda \hat h^2.
\eeq
Thus we obtain the quadratic potential in terms of $\hat h$.
The Planck normalization on the density perturbation (\ref{m}) fixes $\lambda = m^2/2 \simeq 2 \times 10^{-11}$.
Thus, the chaotic inflation with quadratic potential can be realized by the SM Higgs field with the running kinetic term.
Note that all the interaction of the (canonically normalized) Higgs field are suppressed and the system approaches the
free field theory as $h$ increases. 

The BICEP2 result (\ref{B}) has an apparent tension with the Planck data, which can be relaxed by including a large
negative running spectral index \cite{BICEP2}.  A recent work~\cite{Abazajian:2014tqa} have performed a joint analysis 
of the  $Planck$ and BICEP2 datasets, giving a constraint,
\begin{equation}
 \frac{dn_s}{d \ln k} = -0.024 \pm 0.010 ~~  (68 \% {\rm CL}),
\label{eq22}
\end{equation}
which is similar to the combined analysis of $Planck$+WP+highL data~\cite{Ade:2013zuv}.
The analyses assume a scale-independent running, and it is known that  such a large (constant) negative
running would quickly terminate inflation within the e-folding number $30$ or so~\cite{Easther:2006tv}.
This conclusion, however, can be avoided by allowing a scale-dependence of the running, while it remains
more or less constant over the CMB scales.  The simplest way to accomplish this is to add small modulations to the 
inflaton potential~\cite{Kobayashi:2010pz}. The point  is that 
the third derivative of the inflaton potential, which contributes to the running,  can be (locally)
dominated by the modulations, while their contributions to the potential and its first derivative are negligibly small.
Therefore, the running can be enhanced locally without modifying the overall inflaton dynamics. 
In our case,  we can add small modulations to the Higgs potential by introducing another shift symmetry breaking term\footnote{
This additional interactions will contribute
to the Higgs potential in the low energy, but its effect can be absorbed by shifting the values of $\lambda$ and $v$.
}
\bea
{\cal L} &\supset& \Lambda^4 \cos\left(|H|^2/f + \theta\right).
\eea
The precise value of the running depends on the phase of modulations, but
we can evaluate the typical value of the running as
\bea
\left| \frac{dn_s }{ d \ln k} \right| \simeq \left| -2 \frac{V' V'''}{V^2} \right|  \sim \frac{\Lambda^4}{N^{3/2} m^2 f^3},
\eea
where $N$ is the e-folding number.
The running spectral index  (\ref{eq22}) can be realized for e.g. $N=60$, $f \sim 0.1$ 
and $\Lambda^4 \sim m^2 f^2$~\cite{Czerny:2014wua}.

For small field value $h \ll \sqrt{\epsilon}$,
the Lagrangian is reduced to the usual one for the SM Higgs field,
\beq
	\mathcal L \simeq \frac{1}{2}(\partial \tilde h)^2 - \frac{\tilde\lambda}{4}\left(\tilde h^2 - \tilde v^2\right)^2,
\eeq
where  we have defined  $\tilde h \equiv \sqrt{\epsilon} h$,
$\tilde\lambda \equiv \lambda / \epsilon^2$ and $\tilde v \equiv \sqrt{\epsilon}v$. 
In order to explain  the  correct electroweak scale and the 126\,GeV Higgs boson mass,
we must have $\tilde v =246\,$GeV and  $\tilde\lambda \simeq 0.13$.
The SM Yukawa interactions are also obtained in the low energy if we add the Yukawa interactions
with suppressed couplings in \EQ{L}~\cite{Nakayama:2010sk}. That is to say, the low-energy
effective theory coincides with the SM at ${\tilde h} \lesssim \epsilon$, while the theory approaches
the free theory for ${\hat h}$ at ${\tilde h} \gtrsim \epsilon$. The transition from one phase to the other takes
place at the intermediate scale $\sim \GEV{13}$ or above,  whose precise value depends on the running of the
quartic coupling ${\tilde \lambda}$, as shown below.

The quartic coupling ${\tilde \lambda}$ evolves through the renormalization group equations (RGE), and it is known that
 the quartic coupling and its beta function become tiny at high energy scale~\cite{Holthausen:2011aa,Bezrukov:2012sa,Degrassi:2012ry}.
Allowing the RGE running of ${\tilde \lambda}$ in the SM regime, we can estimate
$\epsilon$ as
\bea
\epsilon &\simeq& 1 \times 10^{-5} \sqrt{\frac{{\tilde \lambda}_{\rm IR}}{{\tilde \lambda}_{\rm UV}}} 
\lrfp{\lambda}{2 \times 10^{-11}}{\frac{1}{2}},
\eea
where ${\tilde \lambda}_{\rm UV}$ and ${\tilde \lambda}_{\rm IR}$, respectively, denote the SM 
Higgs quartic coupling evaluated  at the UV and IR energy scales in the scheme of $h < \sqrt{\epsilon}$. 
That is to say, ${\tilde \lambda}_{\rm UV}$ is defined by $\lambda/\epsilon^2$, whereas ${\tilde \lambda}_{\rm IR}$
is fixed by the Higgs boson mass to be ${\tilde \lambda}_{\rm IR} \simeq 0.13$.
Therefore, the smallness of the Higgs quartic coupling at high energy scales, ${\tilde \lambda}_{\rm UV} < {\tilde \lambda}_{\rm IR}$,
is related to the size of $\epsilon$, which parametrizes the explicit breaking of the shift symmetry. 
As ${\tilde \lambda}_{\rm UV} < {\tilde \lambda}_{\rm IR}$ is suggested by the top mass and the Higgs boson mass,
the transition from the SM to the free theory occurs above the intermediate scale, i.e., ${\tilde h} \sim \epsilon \gtrsim \GEV{13}$.


After inflation ends, the Higgs field begins coherent oscillations. As we have seen, while the Yukawa and
gauge interactions of the Higgs field are suppressed at ${\tilde h} \gtrsim \epsilon$, the SM interactions
are reproduced at ${\tilde h} \lesssim \epsilon$.
Thus the particle production during the coherent oscillations is considered to be so efficient that 
the Higgs bosons are thermalized soon.  The reheating temperature therefore tends to be very high, and it could be 
as high as $\sim 10^{13-14}$\,GeV~\cite{Mukaida:2012qn}.
Thermal leptogenesis works for such a high reheating temperature~\cite{Fukugita:1986hr}. 
With the large tensor-to-scalar ratio $r$ given by \EQ{B}, 
the reheating temperature may be probed by the future gravitational wave experiments~\cite{Seto:2003kc,Nakayama:2008wy}.

It is possible to extend our Higgs chaotic inflation model to 
 the  linear or fractional power potential  by imposing a shift symmetry on a certain combination of the Higgs field.
The Higgs chaotic inflation can also be implemented in supergravity~\cite{Takahashi:2010ky,Nakayama:2010kt,Nakayama:2010sk}.
In particular, we can identify the D-flat direction $H_u H_d$ as the inflaton, once we impose a shift symmetry 
as $H_u H_d \to H_u H_d + C$~\cite{Nakayama:2010kt,Nakayama:2010sk}.
In the simplest case, the K\"ahler potential includes a term like, $K \supset (H_uH_d-(H_uH_d)^\dagger)^2$,  in addition to the
usual kinetic terms which break the shifts symmetry,
and the scalar potential of $H_uH_d$ is generated by the superpotential $W = \lambda S H_u H_d$,
where $S$ is the gauge singlet field.\footnote{See Ref.~\cite{BenDayan:2010yz} for a shift symmetry on $H_u$ and $H_d$, instead of on $H_uH_d$.}
The inflaton dynamics is the same as the model of (\ref{L}).
By further imposing $Z_n$ symmetry on $H_uH_d$ with $n$ being an integer, 
Higgs chaotic inflation with fractional power potential can also be realized~\cite{Nakayama:2010sk}.

There is no candidate for dark matter in the SM.
One of the plausible dark matter candidates in the extension of  SM is the QCD axion~\cite{Peccei:1977hh,Kim:1986ax}.
In the present setup, both the inflation scale and the reheating temperature are so  high that  the Peccei-Quinn (PQ) symmetry
is expected to be restored during or after inflation.
As the PQ symmetry gets spontaneously broken  after inflation, the axionic strings as well as domain walls appear.
To avoid the domain wall problem, the domain wall number must be equal to one.
In this case, the axions radiated from the collapse of axionic domain walls gives the dominant contribution to the
relic axion abundance, which requires $f_a \lesssim 3\times 10^{10}$\,GeV where $f_a$ denotes 
the PQ symmetry breaking scale~\cite{Hiramatsu:2012gg}. Another candidate is the sterile neutrino~\cite{Dolgov:2000ew,Boyarsky:2009ix,Kusenko:2009up}, if it is sufficiently light and long-lived. The light mass can be realized by e.g. the 
split seesaw mechanism~\cite{Kusenko:2010ik}, and both the light mass and the longevity can also be explained
by the flavor symmetry~\cite{Ishida:2013mva}. Interestingly, the reheating temperature of order $\GEV{13-14}$ is sufficient 
for producing the right amount of the sterile neutrinos through the $B-L$ gauge boson exchange, 
to account for the observed dark matter density~\cite{Kusenko:2010ik}. At the same time, thermal leptogenesis with two heavy right-handed neutrinos is also possible with this temperature~\cite{Endoh:2002wm,Raidal:2002xf}.
The sterile neutrino dark matter with mass of 
about $7$\,keV~\cite{Ishida:2014dlp,Abazajian:2014gza} can explain
the recent hint for the $3.5$\,keV X-ray line~\cite{Bulbul:2014sua,Boyarsky:2014jta}.

In this letter we have revisited the SM Higgs chaotic inflation in light of the recent discoveries of the SM Higgs at LHC and the primordial B-mode polarization by the BICEP2 experiment, and shown that the quadratic chaotic inflation can be 
realized by the SM Higgs field, based on the running kinetic inflation. 
One of the essential ingredients is the shift symmetry of the Higgs field (\ref{ss}). If 
small modulations to the inflaton potential are  induced by the shift symmetry breaking,  a sizable
running spectral index can also be generated without significant effect on the overall inflaton dynamics. 
If  the primordial B-mode polarization is
more precisely measured by the Planck and other ground-based observations, it will pin down the underlying
inflation model.  Then the next question will be what the inflaton is, which will be important not only for the UV theory but also for 
considering thermal history of the Universe as the baryon asymmetry and dark matter abundance crucially depend on the
reheating temperature in many scenarios. The SM Higgs boson has two important advantages with respect to other candidates for the inflaton. First, it was already discovered and we know that it exists. Second, the reheating takes place through the SM interactions,
and there is no need to introduce additional interactions between the inflaton and the SM sector. As to the second point, one can also
straightforwardly apply the same argument to the $B-L$ Higgs field. Then, the non-thermal leptogenesis will be possible. 
The PQ field can also be the inflaton: it couples to the PQ quarks as well as gluons and hence the reheating takes place successfully.
We also note that the running kinetic inflation can be applied to a gauge singlet inflaton.

\section*{Acknowledgments}
This work was supported by the Grant-in-Aid for Scientific Research on
Innovative Areas (No. 21111006  [KN and FT],  No.23104008 [FT], No.24111702 [FT]),
Scientific Research (A) (No. 22244030 [KN and FT], 21244033 [FT]),  JSPS Grant-in-Aid for
Young Scientists (B) (No.24740135) [FT], and Inoue Foundation for Science [FT].  This work was also
supported by World Premier International Center Initiative (WPI Program), MEXT, Japan.

%
%

\end{document}